\documentclass[preprint,5p]{elsarticle}

\usepackage{amsmath}
\usepackage{amsthm}
\usepackage{amssymb}
\usepackage{bm} 

\usepackage{hyperref}
\usepackage{lineno}
\usepackage{CJK}

\allowdisplaybreaks

\journal{Physics Letter B}

\begin{document}

\begin{CJK*}{UTF8}{}

\begin{frontmatter}

\title{Accurate relativistic density functional for exchange energy of atomic nuclei}

\author[CENS,PKU]{Qiang Zhao}
\author[PKU,JCHP]{Zhengxue Ren}
\author[PKU]{Pengwei Zhao\corref{cor1}}
\ead{pwzhao@pku.edu.cn}
\author[CENS]{Tae-Sun Park}


\cortext[cor1]{Corresponding author}

\address[CENS]{Center for Exotic Nuclear Studies, Institute for Basic Science, Daejeon 34126, South Korea}
\address[PKU]{State Key Laboratory of Nuclear Physics and Technology, School of Physics, Peking University, Beijing 100871, China}
\address[JCHP]{Institut f\"{u}r Kernphysik, Institute for Advanced Simulation and J\"{u}lich Center for Hadron Physics, Forschungszentrum J\"{u}lich, D-52425 J\"{u}lich, Germany}

\begin{abstract}
The inclusion of nucleonic exchange energy has been a long-standing challenge for the relativistic density functional theory (RDFT) in nuclear physics. 
We propose an orbital-dependent relativistic Kohn-Sham density functional theory to incorporate the exchange energy with local Lorentz scalar and vector potentials. 
The relativistic optimized effective potential equations
for the local exchange potentials are derived and solved efficiently. 
The obtained binding energies and charge radii for nuclei  are benchmarked with the results given by the traditional relativistic Hartree-Fock approach, which involves complicated nonlocal potentials. 
It demonstrates that the present framework is not only accurate but also efficient.  
\end{abstract}


\end{frontmatter}


Solving quantum mechanical many-body problems plays an essential role in many fields, such as materials science, condensed matter physics, nuclear physics, etc. 
Density functional theory (DFT) is one of the most successful tools, and no other method achieves comparable accuracy at the same computational costs.
The key task for DFT is to build the \emph{a priori} unknown energy density functional (EDF), whose existence is proved by the Hohenberg-Kohn theorem~\cite{hohenberg1964_PR136_B864} but the exact form is always extremely difficult to determine and must be built with approximations. 
One of the most popular ways is provided by the Kohn-Sham DFT, where the EDF of an interacting system is constructed by introducing an auxiliary noninteracting system moving in a local potential $V_{\rm KS}$ that gives the same ground-state density~\cite{kohn1965_PR140_A1133}.
The EDF for the kinetic energy, the external potential energy, and the Hartree energy can be built straightforwardly, and the so-called exchange-correlation energy functional for the remaining parts is to be addressed.

In nuclear physics, the DFT has been used with great success in describing various phenomena of nuclei throughout the nuclear chart~\cite{bender2003_RMP75_121, Meng2016_book}.
The relativistic density functional theory (RDFT)~\cite{Serot1986Adv.Nucl.Phys.1, ring1996_PiPaNP37_193, meng2006_PiPaNP57_470, vretenar2005_PR409_101} is of particular interest since it exploits basic properties of QCD at low energies, in particular, symmetries and the separation of scales~\cite{Lalazissis2004}. 
By taking into account the Lorentz symmetry, the relativistic density functionals provide an efficient description of nuclei with a delicate interplay between the large Lorentz scalar and vector potentials of the order of hundred MeV, and explain the large spin-orbit splittings and the nuclear magnetic potential in a consistent way~\cite{ring2012_PST150_014035}.  
Over the past decades, a large variety of nuclear phenomena have been described successfully with the RDFT, ranging from infinite nuclear matter to spherical and deformed finite nuclei~\cite{Serot1986Adv.Nucl.Phys.1, ring1996_PiPaNP37_193, meng2006_PiPaNP57_470}, from ground states to collective rotational and vibrational excitations~\cite{vretenar2005_PR409_101,niksic2011_PiPaNP66_519}, from static to dynamic properties~\cite{vretenar2005_PR409_101,ren2020_PLB801_135194, ren2022_PRC105_L011301, ren2022_PRL128_172501}. 

In these studies, the underlying functionals contain only the kinetic and Hartree energies. 
The exchange energies are not taken into account, and the parameters in these functionals are phenomenological and cannot be directly related to realistic nucleon-nucleon interactions.
Such a theoretical framework is relatively simple and the computational costs are low, but it brings following drawbacks: 
(1) the nuclear shell structures and their evolutions are not well reproduced due to the missing of the nucleon-nucleon tensor interactions~\cite{Long2007Phys.Rev.C34314,Shen2018Phys.Lett.B344}; 
(2) an additional adjustment is needed to describe the charge-exchange spin-flip excitations, such as Gamow-Teller and spin-dipole resonances~\cite{liang2008_PRL101_122502,liang2012_PRC85_064302}.

Therefore, it is highly desirable to build a relativistic density functional including the exchange energies within the Kohn-Sham scheme. 
In fact, for Coulombic systems, it often allows the electronic EDFs to depend explicitly on the single-particle orbitals of the Kohn-Sham DFT~\cite{kummel2008_RMP80_3, engel2011_DFTAAC_351}, which are motivated by both practical and formal inadequacies in the conventional functionals, such as the presence of the self-interaction,  the absence of a derivative discontinuity, etc. 
Such orbital-dependent functionals not only include exchange terms, but also provide a clear path toward an \emph{ab initio} DFT. 
Indeed, the total energy can be generally written into orbital-dependent functionals in the many-body perturbation theory, which is directly connected to the microscopic Hamiltonian.

For nuclear systems, the relativistic many-body perturbation theory has achieved great successes in the past decade. 
In its simplest form, the relativistic Hartree-Fock (RHF) calculations reproduce successfully the shell structure evolutions \cite{long2008_EEL82_12001, long2009_PLB680_428, wang2013_PRC87_047301, li2016_PLB753_97, liu2020_PLB806_135524} and spin-isospin resonances \cite{liang2008_PRL101_122502, liang2012_PRC85_064302}.
Moreover, in its resummed form, the relativistic Brueckner-Hartree-Fock (RBHF) theory has a direct connection to the realistic nucleon-nucleon interactions and can  nicely reproduce the binding energies and charge radii of finite nuclei~\cite{Shen2019Prog.Part.Nucl.Phys.103713}.
However, one has to introduce nonlocal potentials in RHF theories, so the theoretical framework is much more involved and the computational costs are extremely heavy~\cite{geng2020Phys.Rev.C101.064302,geng2022Phys.Rev.C105.034329}. 
With contact interactions, the RHF framework can be significantly simplified by representing the exchange terms as Hartree terms via the Fierz transformation~\cite{liang2012_PRC86_021302R,zhao2022_PRC106_034315}, while the exchange of light $\pi$ mesons cannot be considered since it is associated with the long-distance dynamics. 

In this Letter, we construct, for the first time, an orbital-dependent relativistic density functional for the nuclear RHF energy, in which the single-particle RHF orbitals are replaced with the Kohn-Sham orbitals. 
In contrast to the complicated nonlocal potentials involved in the RHF approach, the present orbital-dependent RDFT works in the Kohn-Sham scheme and utilizes fully local relativistic Kohn-Sham (RKS) potentials.
The RKS potentials are determined by the relativistic optimized effective potential (ROEP) method, which is implemented for nuclei for the first time in this work, while it has been widely used in Coulombic systems~\cite{shadwick1989_CPC54_95,engel1995_PRA52_2750,engel1998_PRA58_964,kodderitzsch2008_PRB77_045101a}.
Unlike the Coulombic systems, nuclei are self-bound systems with both spin and isospin degrees of freedom, so the nuclear ROEP equations are strongly coupled in the Lorentz scalar and vector channels. 
Note that the nonrelativistic orbital-dependent DFT has been tested for artificial neutron drops 
based on the simple Minnesota nucleon-nucleon interaction with only central forces~\cite{drut2011_PRC84_014318}. 
Nevertheless, the present density functional is fully relativistic and is based on a well-defined effective Lagrangian including nucleons, mesons, and photons.
It can be used to describe real nuclei and we successfully benchmark our results against the conventional RHF results for nuclei from light to heavy.  

%
The starting point of the present orbital-dependent RDFT is the RHF energy derived from an effective Lagrangian where the nucleons interact with $\sigma$, $\omega$, and $\rho$ mesons as well as the photon.
Following the RHF theory \cite{bouyssy1987_PRC36_380, long2006_PLB640_150}, the RHF energy can be written as 
\begin{align}
	E =& \sum_{a}^{\rm occ.} \langle a| \bm{\alpha\cdot p}+\beta M |a\rangle + \frac{1}{2}\sum_{ab}^{\rm occ.} 
  \langle a b|V(1,2)|b a \rangle \nonumber \\
	&  -\frac{1}{2}\sum_{ab}^{\rm occ.}
  \langle a b|V(1,2)|a b  \rangle, 
  \label{Eq.edf}
\end{align}
where $M$ is the nucleon mass and the two-body interaction $V(1,2)$ includes the following meson- and photon- nucleon interactions:
\begin{subequations}
  \begin{align}
    V_\sigma(1,2) & = -[g_\sigma\gamma^0]_{1}[g_\sigma\gamma^0]_{2} D_\sigma(1,2),  \\
    V_\omega(1,2) & =  [g_\omega\gamma^0\gamma_\mu]_{1}[g_\omega\gamma^0\gamma^\mu]_{2} D_\omega(1,2), \\
    V_\rho(1,2)   & =  [g_\rho\gamma^0\gamma_\mu\vec{\tau}]_{1}\cdot[g_\rho\gamma^0\gamma^\mu\vec{\tau}]_{2} D_\rho(1,2),  \\
    V_A(1,2)      & =  \left[e\gamma^0\gamma_\mu\frac{1-\tau_3}{2}\right]_{1}
                       \left[e\gamma^0\gamma^\mu\frac{1-\tau_3}{2}\right]_{2} D_A(1,2), 
  \end{align}  
\end{subequations}
with $D_\phi$ the propagator for meson and photon fields, $g_\phi$ the coupling constants, $e$ the charge unit, and $\vec\tau$ the isospin Pauli matrices.

The three terms in Eq.~(\ref{Eq.edf}) correspond to the kinetic energy $T$, the Hartree energy $E_{\rm H}$, and the exchange energy $E_{\rm x}$, respectively. 
In contrast to the RHF theory, the occupied single-particle states $|a\rangle$ and $|b\rangle$ in the Kohn-Sham theory should be interpreted as the Kohn-Sham orbitals (defined in Eq. (\ref{equ:RKS_equation})).
As a result, $T$ and $E_{\rm x}$ are orbital-dependent functionals, while $E_{\rm H}$ can be written explicitly as a functional of the scalar density $\rho_{s}$ and the vector currents $j^\mu$ that are defined with the Kohn-Sham orbitals in the coordinate space by, 
\begin{align}
  \rho_{s,\tau}(\bm r)
  &=\sum_{a\in \tau}^{\rm occ.}
    \bar\varphi_{a}(\bm r)
    \varphi_{a}(\bm r), \\
  j^\mu_\tau(\bm r)
  &=\sum_{a\in \tau}^{\rm occ.}
    \bar\varphi_{a}(\bm r)
    \gamma^\mu
    \varphi_{a}(\bm r).
\end{align}
Here, the subscript $\tau$ is used to distinguish neutron and proton.
 
According to the Kohn-Sham DFT, the RKS orbitals should be calculated by solving the RKS equation, which is essentially a Dirac equation,
\begin{align}\label{equ:RKS_equation}
  \left\{-i\bm{\alpha\cdot\nabla}
      +\beta\left[
      M+S_\tau(\bm r)+\gamma_\mu V^\mu_\tau(\bm r)
      \right]\right\}
\varphi_{a}(\bm r)
=\varepsilon_{a}\varphi_{a}(\bm r),
\end{align}
in which $S_{\tau}$ and $V_{\tau}^\mu$ are the RKS potentials defined via the variation of the energy functional,
\begin{align}
    S_\tau (\bm r)&
  = S_{\rm H,\tau} (\bm r) + S_{\rm x,\tau}(\bm r)
  =\frac{\delta E_{\rm H}}{\delta \rho_{s,\tau}}
  +\frac{\delta E_{\rm x}}{\delta \rho_{s,\tau}}, \\
  V_\tau^\mu (\bm r)&
  = V^\mu_{\rm H,\tau}(\bm r) + V^\mu_{\rm x,\tau}(\bm r)
  =\frac{\delta E_{\rm H}}{\delta (j_\tau)_\mu}
  +\frac{\delta E_{\rm x}}{\delta (j_\tau)_\mu}.
  \end{align}
Each potential is local and is decomposed into the Hartree and exchange parts.
Since the Hartree energy $E_{\rm H}$ is an explicit functional of density and currents, one can readily obtain the Hartree potentials $S_{\rm H,\tau} (\bm r)$
and $V^\mu_{\rm H,\tau}(\bm r)$ by variation. 
However, it is nontrivial to obtain the exchange potentials $S_{\rm x,\tau} (\bm r)$
and $V^\mu_{\rm x,\tau}(\bm r)$ because the exchange energy $E_{\rm x}$ explicitly depends on the Kohn-Sham orbitals rather than the density or currents.   

The general procedure to get the exchange RKS potentials is to solve the ROEP equations, which are derived here via the chain rule of functional differentiation and the details can be seen in the Supplemental Material~\cite{supp}. 
The nuclear ROEP equations consist of a pair of coupled equations for the scalar and vector potentials, 
\begin{subequations}
  \begin{align}
    &\sum_{a\in\tau}^{\rm occ.}
    \big[
      \bar\xi_{a}(\bm r)\varphi_{a}(\bm r)
      + {\rm c.c.}
    \big] = 0,
    \label{equ:ROEP2_s} \\
    &\sum_{a\in\tau}^{\rm occ.}
    \big[
      \bar\xi_{a}(\bm r)\gamma^\mu\varphi_{a}(\bm r)
      +{\rm c.c.}
    \big] = 0.
    \label{equ:ROEP2_v}
  \end{align}
\end{subequations}
Here, $\bar\xi_a$ are the first-order changes in the RKS orbital $\bar\varphi_a$ when the potential $S_{\rm x, \tau} +\gamma_\mu V^\mu_{\rm x, \tau}$ in the RKS equation (\ref{equ:RKS_equation}) is replaced with the orbital-specific potential $W_a(\bm r) = \frac{1}{\bar\varphi_a }\frac{\delta E_{\rm x}}{\delta \varphi_a}$.
In this sense, the ROEP equations (\ref{equ:ROEP2_s}) and (\ref{equ:ROEP2_v}) are consequences of the vanishing first-order corrections for the density and currents. 
By introducing the first-order correction $\zeta_{a}$ of the eigenvalue $\varepsilon_a$, the ROEP equations can also be written equivalently as 
\begin{align}
  S_{\rm x,\tau} \rho_{s,\tau} &+ V^\nu_{\rm x,\tau} (j_{\tau})_\nu \nonumber \\
  &=\frac{1}{2}\sum_{a\in\tau}^{\rm occ.}
    \bigg\{ \bar\varphi_a W_a \varphi_{a}
    + T_{\rm es1} + T_{\rm os1} + {\rm c.c.}
    \bigg\},
    \label{equ:ROEP1} \\
  S_{\rm x,\tau}j^\mu_{\tau} &+ V^\mu_{\rm x,\tau} \rho_{s,\tau} \nonumber \\
  &=\frac{1}{2}\sum_{a\in\tau}^{\rm occ.}
    \bigg\{
      \bar\varphi_a W_a \gamma^\mu\varphi_{a}
      + T_{\rm es2}^\mu + T_{\rm os2}^\mu + {\rm c.c.}
    \bigg\},
    \label{equ:ROEP2} 
\end{align}
in which $T_{\rm es1}$ and $T_{\rm es2}^\mu$ are the energy-shift terms
\begin{align}
  T_{\rm es1}
  &= -\zeta_a\bar\varphi_a\gamma^0\varphi_a, \\
  T_{\rm es2}^\mu
  &= -\zeta_a\bar\varphi_a\gamma^0\gamma^\mu\varphi_a, 
\end{align}
$T_{\rm os1}$ and $T_{\rm os2}^\mu$ are the orbital-shift terms
\begin{align}    
  T_{\rm os1}
  &= \bar\xi_{a} 
  \left[
    i\bm\gamma\cdot\overleftarrow{\bm\nabla} - \gamma^0 \varepsilon_{a}
  \right]\varphi_{a}, \\
  T_{\rm os2}^\mu
  &=\bar\xi_{a}
  \left[
    i\bm\gamma\cdot\overleftarrow{\bm\nabla}
    -\gamma^0 \varepsilon_{a}
    +\gamma_\nu V_\tau^\nu 
  \right]
  \gamma^\mu\varphi_{a}.
\end{align}
Since $\zeta_{a}$ and $\bar\xi_{a}$ correspond to the first-order perturbation of the energy $\varepsilon_a$ and the RKS orbital, respectively, 
the energy-shift and orbital-shift terms should be smaller than the $W_a$ terms on the righthand sides of Eqs. (\ref{equ:ROEP1}) and (\ref{equ:ROEP2}). 

As a demonstration, the present framework is applied to the spherical nuclei $^{16}$O, $^{40}$Ca, $^{48}$Ca, $^{132}$Sn, and $^{208}$Pb.
The calculated results are benchmarked with the RHF results with the widely used effective interaction PKO2~\cite{long2008_EEL82_12001}.
We start with an initial guess for the RKS potentials, and solve the RKS equation by expanding the RKS orbitals in a set of spherical Dirac Woods-Saxon (DWS) basis~\cite{zhou2003_PRC68_034323}, which is constructed in a radial box with the size $R_{\rm box}=16$ fm and the mesh $\Delta r=0.1$ fm.
The RKS equations provides the RKS orbitals, which are used to calculate the scalar and vector densities, and in turn, the Hartree potentials. 
Then, the ROEP equations are solved to obtain the exchange parts of the RKS potentials. 
Note that the orbit-shift terms are always neglected here since they are quite small compared to other terms. 
This is consistent with the widely used relativistic Krieger-Li-Iafrate (RKLI) approximation~\cite{kreibich1998_PRA57_138, krieger1990_PLA146_256} for Coulombic systems, while the validity of this approximation is confirmed for nuclear systems in the present work.
Updating the Hartree and exchange potentials, the RKS equation is solved iteratively until convergence is achieved. 


\begin{figure}[!htbp]
  \centering
  \includegraphics[width=0.44\textwidth]{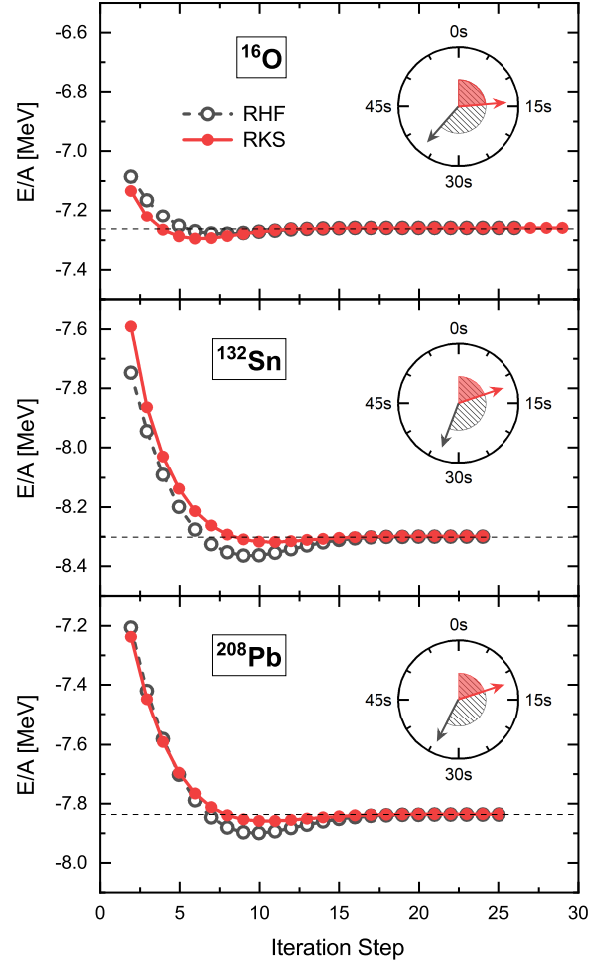}
  \caption{(Color online).
  Total energy per nucleon as a function of the iteration step for $^{16}$O, $^{132}$Sn, and $^{208}$Pb. The inset clocks represent the total computing time to get convergence with the RKS (red) and RHF (black) methods. All calculations are performed on a desktop computer.
  }
  \label{fig:iteration}
\end{figure}

In Fig. \ref{fig:iteration}, our results for the total energy per nucleon as a function of the iteration steps for $^{16}$O, $^{132}$Sn, and $^{208}$Pb are compared with the corresponding RHF results.
During the iteration, the calculated energies with both methods smoothly converge to almost the same value with a similar convergence pattern. 
This clearly demonstrates the validity of the present orbital-dependent RDFT framework. 

Although the iteration numbers to achieve convergence for the present RKS calculations are almost the same as the RHF calculations, the total computing time is much less; only about one third of the computing time of the RHF calculations, as can be seen in the inset clocks. 
Note that the RHF equation is a complex integro-differential equation with nonlocal exchange potentials, while all the potentials in the present RKS equation are local. 
As a result, it is not necessary to calculate the matrix elements of the nonlocal potential on the DWS basis in the solution of the RKS equation.
This is the main reason of the computational merits, although this advantage comes at the price of the additional solution of the ROEP equations.
In the present work, the ROEP equations are solved by neglecting the insignificant orbit-shift terms, but a preliminary full solution of the ROEP equations with the orbital-shift terms shows that the computing time does not change significantly.
Note that a high accuracy full solution of the ROEP equations is quite a tricky problem, which has been discussed extensively in the field of Coulombic calculations \cite{kodderitzsch2008_PRB77_045101a, jiang2005.JCP123.224102}.
Works along this direction are in progress.


\begin{figure}[!htbp]
  \centering
  \includegraphics[width=0.44\textwidth]{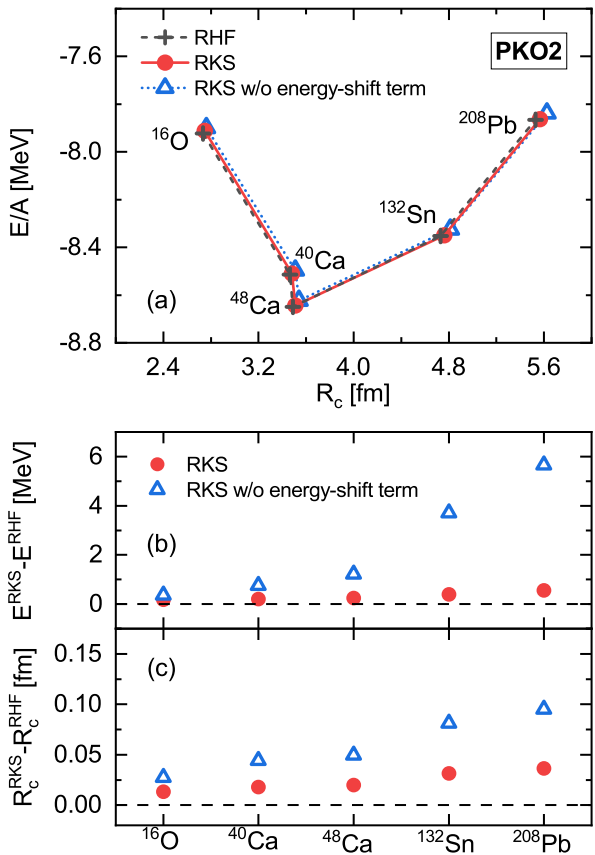}
  \caption{(Color online).
  (a) Total energies per nucleon as a function of the charge radii for selected spherical nuclei. The crosses, solid  circles, and open triangles denote the RHF results, the RKS results, and the RKS results neglecting the energy-shift terms in the ROEP equations, respectively. Differences in the total energies (b) and the charge radii (c) between the RKS and RHF results with and without the energy-shift term are also shown. Note that the microscopic center-of-mass correction energies~\cite{bender2000_EA7_467, long2004_PRC69_034319, zhao2009_CPL26_112102} are included here. 
  }
  \label{fig:BE-Rc}
\end{figure}

The influence of the orbit-shift and energy-shift terms on the calculated total energies and charge radii is shown in detail in Fig.~\ref{fig:BE-Rc}.
The ground-state binding energies and charge radii of $^{16}$O, $^{40}$Ca, $^{48}$Ca, $^{132}$Sn, and $^{208}$Pb are calculated within the present RKS framework, and are compared with the RHF results.  
Note that the microscopic center-of-mass correction~\cite{bender2000_EA7_467, long2004_PRC69_034319, zhao2009_CPL26_112102} is considered in both RKS and RHF calculations after the iteration converges.
As can be seen in Fig.~\ref{fig:BE-Rc}(a), the RHF results are well reproduced by the present RKS calculations where the orbit-shift terms are neglected.
By further neglecting the energy-shift terms, the RKS results are still in a reasonable agreement with the RHF results, and visible deviations can be seen only for heavy nuclei such as $^{132}$Sn and $^{208}$Pb.
This demonstrates that the $W_a$ terms in the ROEP equations alone provide a good estimation for the local exchange potentials.

A more detailed comparison is made in Figs.~\ref{fig:BE-Rc}(b) and \ref{fig:BE-Rc}(c) by depicting the differences in the total energies and charge radii between the RKS and RHF results
with and without the energy-shift term.
Although both the RKS and RHF calculations aim to minimize the RHF energy, the former should be regarded as a constrained optimization of the total energy by requiring local auxiliary potentials only.
Therefore, the ground-state energies obtained via RKS are expected to be higher than those of RHF, and this is consistent with our calculations as shown in Fig. \ref{fig:BE-Rc}(b). 
The small differences between the RKS and RHF results demonstrate that the orbit-shift terms can be safely neglected for nuclear systems. 
The energy-shift terms, however, are relatively more important in particular for heavy nuclei. 
For example, the energy difference between the RKS and RHF results can reach to 5.7 MeV for $^{208}$Pb.

\begin{figure}[!htbp]
  \centering
  \includegraphics[width=0.44\textwidth]{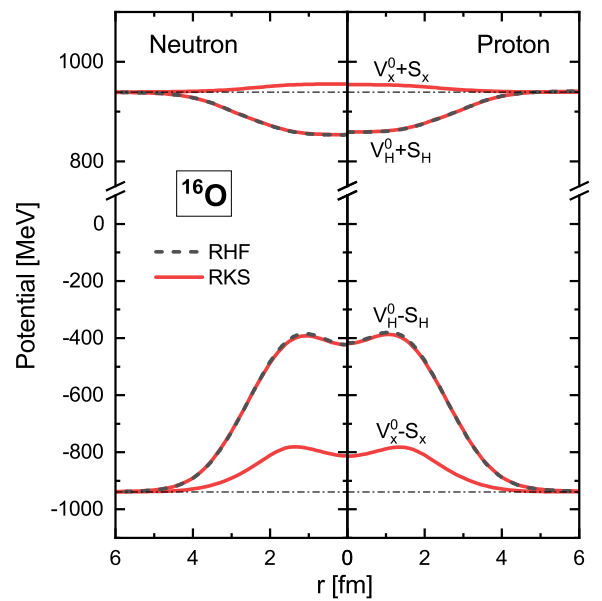}
  \caption{(Color online).
  The relativistic Kohn-Sham potentials for $^{16}$O in comparison with the RHF mean potentials.
  }
  \label{fig:potential}
\end{figure}

The main difference between the present RKS and the RHF approaches is that the latter involves nonlocal potentials, whereas the former employs only local Kohn-Sham potentials. 
In Fig.~\ref{fig:potential}, the nucleon and anti-nucleon RKS potentials for $^{16}$O are depicted in comparison with the RHF mean potentials.
The Hartree potentials are local for both RKS and RHF approaches, and they nicely agree with each other. 
The magnitudes for the Hartree potentials are around several tens MeV in the Fermi sea and several hundreds MeV in the Dirac sea, which is associated with the large spin-orbit splittings in nuclei. 
The exchange potentials in the RHF method are nonlocal and cannot be directly compared to local potentials. 
In the present RKS framework, however, the exchange potentials are fully local. 
It is seen that the magnitudes of the exchange potentials are smaller than those of the Hartree potentials in both the Fermi sea and Dirac sea. 
In particular, in the Fermi sea, the obtained exchange potential is slightly repulsive in contrast to the attractive Hartree potential. 
As a result, they provide opposite contributions to the total energy.

\begin{figure}[!htbp]
  \centering
  \includegraphics[width=0.44\textwidth]{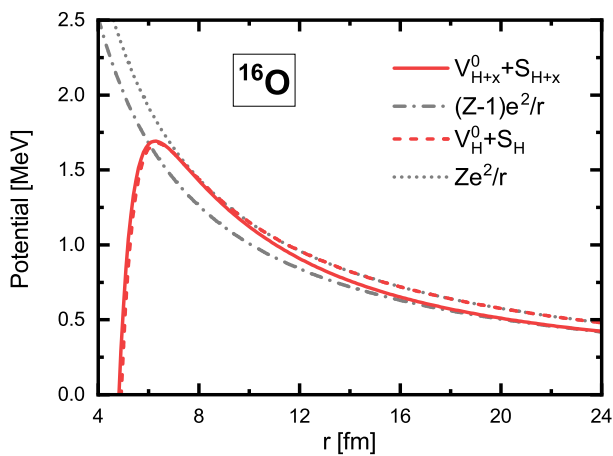}
  \caption{(Color online).
  The proton Kohn-Sham potentials for $^{16}$O at large radial distances.
  }
  \label{fig:coulomb}
\end{figure}

It is of importance to have the correct long-range behavior of the Coulomb potential for protons, i.e., $\sim (Z-1)e^2/r$ ($Z$ is the proton number).
This is in fact not a trivial condition and the Hartree potential decays as $\sim Ze^2/r$ due to the self-interaction.
Figure \ref{fig:coulomb} shows the proton potentials of $^{16}$O at large radial distances given by the RKS calculation.
By including the exchange energy that removes the self-interaction, we are pleased to find that the RKS calculation nicely reproduces the correct long-range behavior.


In summary, an orbital-dependent relativistic energy density functional including the nucleonic exchange energy has been constructed for the first time within the relativistic Kohn-Sham framework for the nuclear RHF energy. 
The single-particle RHF orbitals are replaced with the Kohn-Sham orbitals, which are obtained by solving the relativistic Kohn-Sham equation with local Lorentz scalar and vector potentials. 
Although it requires an additional solution of the ROEP equations to obtain the local exchange potentials, the present framework is superior to the traditional RHF approach which involves complex nonlocal potentials. 
Benchmark calculations have been performed for the ground-state energies and charge radii of nuclei from light to heavy, and it demonstrates that the proposed framework is not only highly accurate but also efficient.  

Since the exchange potentials are local, this new framework could be straightforwardly extended to nuclei with arbitrary deformation by solving the RKS equations on three-dimensional lattice~\cite{ren2017Phys.Rev.C95.024313,li2020Phys.Rev.C102.044307}. 
Moreover, it provides a promising way to construct an \emph{ab initio} relativistic energy density functional for atomic nuclei based on realistic nucleon-nucleon interactions, such as the Bonn potentials~\cite{machleidt1989_AiNP_189} and the relativistic chiral interactions~\cite{ren2018_CPC42_014103, lu2022.PRL128.142002}, softened with the Brueckner G-matrix approach~\cite{shen2016_CPL33_102103,shen2017_PRC96_014316} and/or modern renormalization group methods~\cite{bogner2010.PiPaNP65.94}.

\section*{Acknowledgments}
The authors thank J. Geng and W.H. Long for helpful discussions.
This work is supported by the Institute for Basic Science (Grant No. IBS-R031-D1), 
the National Key R\&D Program of China (Contracts No. 2017YFE0116700 and No. 2018YFA0404400), 
and the National Natural Science Foundation of China (Grants No. 11875075, No. 11935003, No.11975031, and No. 12141501).
Z.R. is supported in part by the European Research Council (ERC) under the European Union's Horizon 2020 research and innovation programme (Grant agreement No. 101018170).

\appendix
\section{Supplementary material}
\label{appendix}
Supplementary material related to this article can be found online at \url{url}.

\clearpage
\end{CJK*}

\end{document}